\documentclass[letterpaper,10pt]{article} 
\usepackage{amsmath,amssymb,amscd,latexsym,dsfont}
\usepackage{comment}
\usepackage{color}
\usepackage{subfigure}
\usepackage{enumerate}
\usepackage{enumitem}
\usepackage{stfloats}
\usepackage{psfrag} 
\usepackage{setspace}
\usepackage{cite}
\usepackage{balance}
\usepackage{enumitem}
\usepackage{tikz}
\usepackage{pgfplots,amsfonts}
\usepackage{pgfplotstable}
\usetikzlibrary{shapes}
\usetikzlibrary{spy}
\usetikzlibrary{fit}
\usetikzlibrary{shapes.multipart}
\usetikzlibrary{positioning}
\pgfplotsset{compat=1.14}
\usepackage{wrapfig}
\usepackage[acronym,nomain]{glossaries}
\usepackage{arydshln}
\usepackage{wrapfig}

\usepgfplotslibrary{fillbetween}
\usepackage{tikz-network} 
\usetikzlibrary{arrows.meta} 
\usetikzlibrary{calc}
\usepackage{colortbl} 
\usetikzlibrary{backgrounds} 
\usepackage{enumerate}
\usepackage{opticameet3} 
\newcommand\authormark[1]{\textsuperscript{#1}}
\usepackage{bm}
\usepackage{soul}
\usepackage{amsmath,amssymb}
\usepackage[colorlinks=true,bookmarks=false,citecolor=blue,urlcolor=blue]{hyperref} 
\usepackage{pifont}
\begin{document}

\title{Predicting Nonlinear Interference for Short-Blocklength 4D Probabilistic Shaping}

\copyrightyear{2024}

\author{Jingxin Deng\authormark{1}, Bin Chen\authormark{1,*}, Zhiwei Liang\authormark{1}, Yi Lei\authormark{1} and Gabriele Liga\authormark{2}
}

\address{\authormark{1} School of Computer Science and Information Engineering, Hefei University of Technology, Hefei, China\\
\authormark{2} Department of Electrical Engineering, Eindhoven University of Technology, Eindhoven, The Netherlands\\
}

\email{\authormark{*}bin.chen@hfut.edu.cn}

\begin{abstract}
We derive a heuristic nonlinear interference model for 4D probabilistic shaping considering the polarization and time correlation of the 4D symbols. We demonstrate an average SNR prediction gap from split-step Fourier simulations of 0.15~dB. 
\end{abstract}

\section{Introduction}
Probabilistic shaping (PS) has demonstrated excellent performance in the additive white Gaussian noise (AWGN) channel and optical systems. When dealing with nonlinear interference (NLI) in optical fiber transmission, conventional PS may be more susceptible to performance degradation \cite{8895789}. 
Recently, multi-dimensional constellations have been shown to have greater shaping gains compared to traditional polarization-multiplexed two-dimensional (PM-2D) formats in optical fiber channels \cite{7839914}. Therefore, several works aimed to improve the NLI performance by designing multi-dimensional PS schemes with short blocklength \cite{9125832,9605980}, which can lead to higher signal-to-noise ratio (SNR) at the receiver. 
In order to find a nonlinearity-tolerant shaped constellation in the nonlinear fiber channel, powerful analytical models are crucial for quickly and accurately predicting the induced NLI.

To estimate modulation format dependent NLI in nonlinear fiber propagation, 
the enhanced Gaussian noise (EGN) model \cite{Carena:14} is among the most popular analytical models. However, the EGN model is only valid for PM systems that independently transmit 2D formats over the two orthogonal polarizations.  
New analytical models \cite{GabrieleEntropy2020,10255078} have recently been proposed to predict the impact of NLI on general dual-polarization 4D (DP-4D) modulation, where the symbols are jointly mapped over the two polarizations of the optical field. Like the EGN model, this 4D model operates under the assumption of i.i.d. input symbols, which may not be accurate for short blocklength shaping schemes, where temporal symbol correlations must be taken into account. 
To capture the effect of correlated symbols in time domain on the NLI, heuristic approaches such as the finite-memory GN model \cite{6824164} and the so-called energy dispersion index \cite{9464637},  have been proposed. More recently, in \cite{9716753}, a heuristic model introducing the idea of a windowed kurtosis within the EGN model framework was proposed.
However, an accurate NLI model able to capture the temporal correlations arising in 4D-PS schemes with finite blocklength has yet to be devised.

In this work, we extend the approach in \cite{9716753} to  DP-4D PS formats and derive a new windowed 4D (W-4D) NLI model by introducing time-window averaged moments of the 4D constellation into the model in \cite{10255078}. Our preliminary findings show that the resulting heuristic model can effectively account for both the polarization and time correlations of the 4D symbols. 
The proposed W-4D model shows good agreement with the split-step Fourier method (SSFM) simulation results, with SNR prediction error within 0.15~dB for all the considered 4D-PS formats with finite blocklength.  Moreover, we show that the existing 2D and 4D models may lead to inaccuracies in the prediction of the SNR of up to 1.72~dB.

\vspace{-0.6em}
\section{System Model and The Proposed Windowed 4D NLI Model}
\vspace{-0.2em}
\subsection{System Model}
\begin{figure}[!tb]
\vspace{-1.6em}
  \centering
  \includegraphics[width=\textwidth]{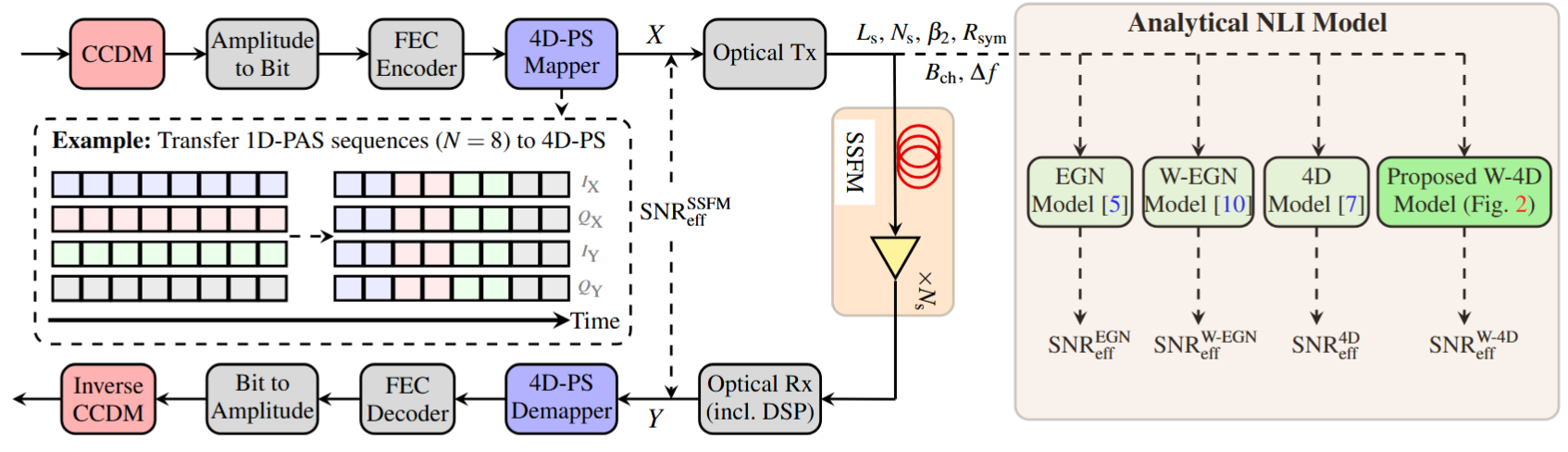}
    \vspace{-1.5em}
    \caption{Block diagram of 
    optical transmission system with 4D-PS signal. Left: End-to-end implementation with 4D-PS encoder and decoder. The 4D-PS mapper is implemented to transfer 1D-PAS to 4D-PS. 
    Right: SNR prediction of the considered four kinds of analytical NLI models.}
    \label{system model}
    \vspace{-2.2em}
\end{figure}

Fig.~\ref{system model} shows the key blocks for 4D-PS signaling and the NLI models analyzed in this work. 
At the transmitter side, a sequence of i.i.d. bits is shaped using CCDM as 1D probabilistic amplitude shaping (PAS), which converts the information bits into amplitude codewords. The amplitudes are then labeled back into bits and fed into an forward error correction (FEC) encoder.
For the considered optical transmission system, there are four dimensions available denoted as $I_\text{x}$, $Q_\text{x}$, $I_\text{y}$ and $Q_\text{y}$, representing the two quadrature components in each polarization dimension, respectively. Extending the time dependency of 1D symbols to four dimensions of dual polarization is achieved using a 4D-PS mapper (see the example in Fig.~\ref{system model}) to generate 4D real symbols, which are then transmitted through the optical fiber channel. Note that this is only one of the possible approaches for generating 4D-PS symbols. 
At the receiver, digital signal processing (DSP) is used, including chromatic dispersion compensation (CDC), matched filter and sampling. 
 The effective SNR via SSFM (denoted by $\mathrm{SNR}_{\mathrm{eff}}^{\text {SSFM }}$) is obtained using the transmitted symbols $X$ and the received symbols $Y$. Then the output symbols $Y$ are processed by a 4D-PS demapper where soft information is computed and passed to an FEC decoder. Finally, inverse CCDM is performed to recover the transmitted bits. 

The analytical models for estimating NLI power and SNR are shown on the right of Fig.~\ref{system model}. The models are used to replace the time-consuming SSFM simulations in order to predict NLI and effective SNR.
The NLI power is a function of the transmitted signal and the parameters of the fiber link, including symbol rate $R_\text {sym }$, channel bandwidth $B_{\text{ch}}$,  fiber dispersion coefficient $\beta_{2}$,  span length $L_{\text{s}}$, number of spans $N_{\text{s}}$ and channel spacing $\Delta f$.
Thus, given the parameters, effective SNRs can be estimated via different NLI models, including EGN model, windowed EGN model, 4D model, and the proposed windowed 4D model.

\vspace{-0.6em}
\subsection{The  Proposed Windowed 4D NLI Model}
\vspace{-0.2em}

\begin{figure}[!b]
 \vspace{-1.2em}
    \centering
    \includegraphics[width=0.99\textwidth]{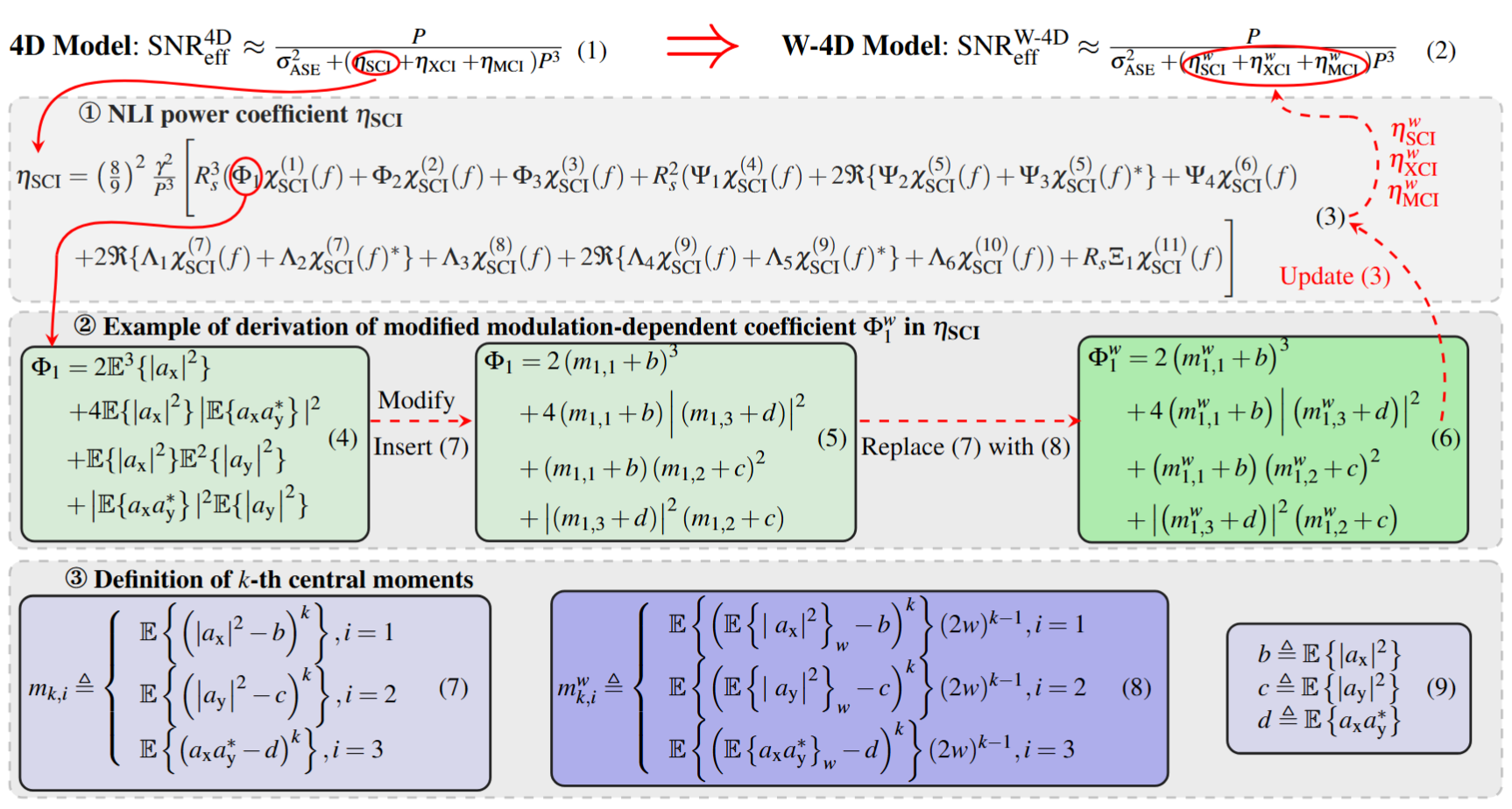}
    \vspace{-1em}
    \caption{The derivation  of $\mathrm{{SNR}_{\text {eff }}^{\text{W-4D}}}$ for W-4D NLI model.}
    \label{formats}
       \vspace{-2.6em}
\end{figure}

In order to improve the accuracy of the effective SNR prediction for 4D-PS with finite blocklength.
Fig.~\ref{formats} shows the derivation of $\mathrm{SNR}_{\mathrm{eff}}^{\text {W-4D}}$ for the proposed W-4D model, using a first-order perturbation approach under the hypothesis of wavelength division multiplexed (WDM) transmission. This derivation extends the 4D model in \cite{10255078} introducing  the symbol energy correlations within a sliding window as done in \cite{9716753} for the simple EGN model.

As shown in Fig.~\ref{formats}, the effective SNR can be calculated as (1) by the NLI power coefficients ($\eta_{\text{SCI}}$, $\eta_{\text{XCI}}$, $\eta_{\text{MCI}}$) via the 4D model which was only suitable for long blocklength constellations. For short blocklength shaping constellations, the 4D model was extended to the W-4D model (see (2)). We show the derivation of the W-4D model as following. 
For simplicity, we only take the $\eta_{\text{SCI}}$ (see the gray block~$\textbf{{\ding{172}}}$) as an example to show the derivation process. 
The $\eta_{\text {XCI}}$ and $\eta_{\text {MCI}}$ were derived following an approach similar to $\eta_{\text {SCI}}$.  
In (3),  $\chi_\text{SCI}^{(i)}, i=1,2,...,11$ are integrals related to channel parameters. The modulation-dependent NLI coefficients $\Phi_1$, $\Phi_2$, $\Phi_3$, $\Psi_1$, ..., $\Psi_4$, $\Lambda_1$, ..., $\Lambda_6$, $\Xi_1$ are functions of the transmitted modulation formats with independent symbols and are given in \cite[Table~2]{10255078}. For the shaping constellations with dependent symbols, the modulation-dependent NLI coefficients need to calculate with a time-window averaged moments of the shaping constellations (see the gray block~$\textbf{{\ding{173}}, \ding{174}}$). 
Taking $\Phi_1$ as an example, $\Phi_1$ (5) can be obtained by modifying (4) by inserting (7). 
And $\Phi_1^w$ (6) with a moving average filter with window length $w$ can be obtained by replacing $m_{k,i}$ (7) with $m_{k,i}^w$ (8).  
Then, the $\mathrm{SNR}_{\mathrm{eff}}^{\text {W-4D}}$ can be obtained by using the modified NLI power coefficients $\eta_{\text {SCI }}^{\text{w}}$,  $\eta_{\text {XCI }}^{\text{w}}$ and $\eta_{\text {MCI }}^{\text{w}}$.

\vspace{-0.6em}
\section{Numerical Results}
\vspace{-0.5em}
In this section, we first analyze the accuracy of the EGN, windowed EGN, 4D NLI and W-4D NLI models, where  4D-PS is implemented with CCDM based on 4D mapper shown in Fig.~\ref{system model}. 
The multi-span optical fiber link we considered in this work includes 9 WDM channels each with 45~GBaud symbol rate. The WDM channel spacing is 50~GHz, and a root-raised-cosine filter with a roll-off factor of 0.01$\%$ is used for pulse shaping. The parameters of the fiber link are as follows: attenuation coefficient $\alpha$ = 0.2~dB/km, dispersion parameter $D$=17~ps/nm/km and nonlinear coefficient $\gamma$ = 1.3~/W/km. Each span consists of an 80~km single-mode fiber followed by an EDFA with a noise figure of 5~dB, and the total distance is 320~km.

\begin{figure}[!tb]
 \vspace{-1.75em}
\centering
 \includegraphics[width=0.992\textwidth]{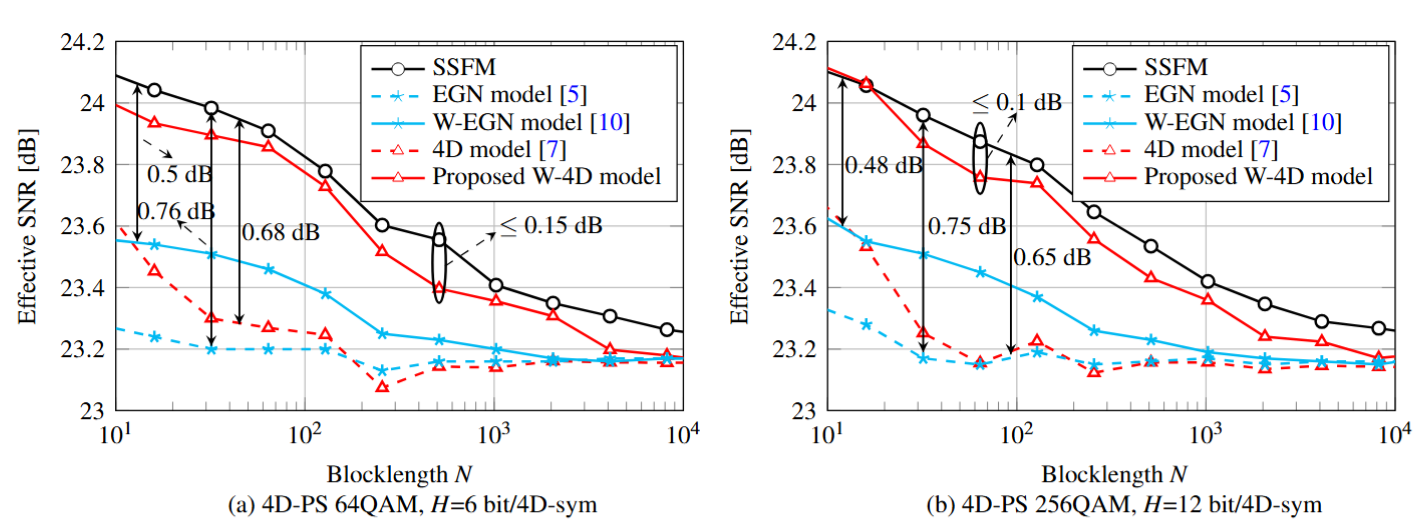}
 \vspace{-1.4em}
  \caption{Effective SNR vs. blocklength $N$ for nonlinear fiber channel at optimal launch power.} 
  \label{PS-64QAM H=6 and PS-256QAM H=12}
   \vspace{-2.2em}
\end{figure}

In Figs.~\ref{PS-64QAM H=6 and PS-256QAM H=12} (a) and (b), the effective SNRs are evaluated by using different models and modulation formats for nonlinear fiber channel at optimal launch power.  It shows that only the proposed W-4D model can accurately predict SNR, and the maximum deviation about 0.15~dB for all blocklengths, while the maximum underestimate error of the EGN model, W-EGN model and 4D model reaches 0.76~dB, 0.5~dB and 0.68~dB, respectively. As the blocklength increases, the prediction deviation of the W-4D model, along with other models, decreases progressively. For the blocklength $N=10000$, all the considered models can guarantee an accurate prediction of the SNR 
(an average gap about 0.1~dB).

\begin{wrapfigure}{r}{0.425\textwidth}
\vspace{-2.1em}
\centering
 \includegraphics[width=0.44\textwidth]{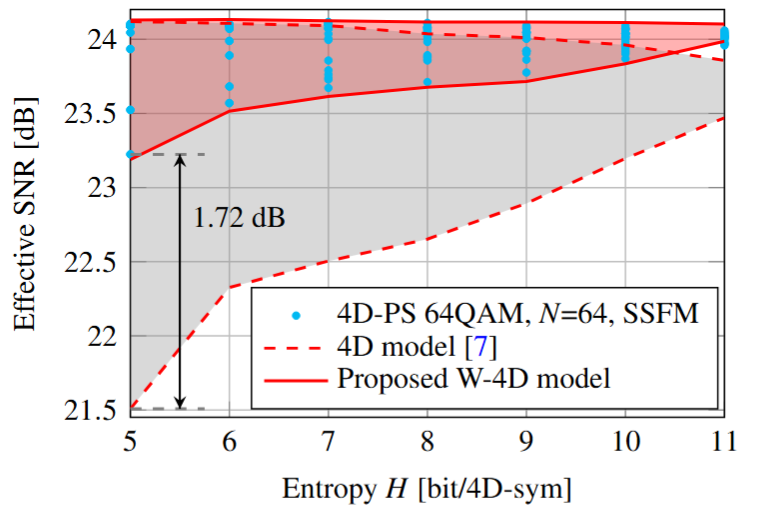}
\vspace{-2.5em}
\caption{Effective SNR vs. entropy $H$ of 4D-PS 64QAM.}
\label{fig:result2}
\vspace{-2em}
\end{wrapfigure}

In order to further evaluate the accuracy of SNR prediction for different 4D probability distributions, 4D-PS 64QAM are generated for different entropies by using 1D PAS sequences.  
For example, for PS-64QAM with 1D probability distribution $P_{1\text{D}}=[0.4,0.3,0.2,0.1]$, we take the permutations of these four probabilities, resulting in 24 4D-PS  distributions for the same entropy that are not factorizable on the 4 individual dimensions. 
We evaluated the prediction accuracy of the 4D model and the W-4D model by estimating the maximum and minimum effective SNR for 4D-PS 64QAM with various entropies, respectively. The simulation results are depicted in Fig.~\ref{fig:result2}. For different entropies, the proposed W-4D model can accurately predict the SNR, while the maximum deviation of SNR prediction using the previous 4D NLI model is  up to  1.72~dB.

\vspace{-0.6em}
\section{Conclusions}
\vspace{-0.5em}
In this paper, we proposed a W-4D NLI model to predict the NLI noise for 4D-PS by considering the effect of correlated symbols in both the dual polarization and time domain.
 Our results show that for different 4D-PS constellations with various 4D probability distributions, the proposed W-4D NLI model can ensure an average gap within 0.15~dB  compared to SSFM, while the error of the previous 4D NLI model is up to 1.72~dB.

\begin{footnotesize}
\vspace{0.1em}
\textbf{Acknowledgements:} {The work of B.~Chen and  Y.~Lei are supported by the National Natural Science Foundation of China (NSFC)
under Grant 62171175 and State Key Laboratory of Advanced Optical Communication Systems and Networks, China.
\par}
\end{footnotesize}

\end{document}